# Optical control of carrier wavefunction in magnetic quantum dots

P. Zhang[1], T. Norden[1], J. M. Pientka[2], R. Oszwałdowski[3], A. Najafi[1], B. Barman[4], Y. Tsai[1], W-C. Fan[5], W-C. Chou[5], J. E. Han[1], I. Žutić[1], B. D. McCombe[1] and A. Petrou[1]

[1] *Department of Physics, University at Buffalo SUNY, Buffalo, NY 14260 USA*

[2] *Department of Physics, St. Bonaventure University, St. Bonaventure, NY 14778 USA*

[3] *Department of Physics, South Dakota School of Mines & Technology, Rapid City, SD 57701 USA*

[4] *Department of Physics, University of Michigan-Flint, Flint, MI 48502 USA*

[5] *Department of Electro-physics, National Chiao Tung University, Hsinchu 300, Taiwan*

**ABSTRACT**

Spatially indirect Type-II band alignment in magnetically-doped quantum dot (QD) structures provides unexplored opportunities to control the magnetic interaction between carrier wavefunction in the QD and magnetic impurities. Unlike the extensively studied, spatially direct, QDs with Type-I band alignment where both electrons and holes are confined in the QD, in ZnTe QDs embedded in a (Zn,Mn)Se matrix only the holes are confined in the QDs. Photoexcitation with photon energy 3.06 eV (2.54 eV) generates electron-hole pairs predominantly in the (Zn,Mn)Se matrix (ZnTe QDs). The photoluminescence (PL) at 7 K in the presence of an external magnetic field exhibits an up to three-fold increase in the saturation red shift with the 2.54 eV excitation compared to the shift observed with 3.06 eV excitation. This unexpected result is attributed to multiple hole occupancy of the QD and the resulting increased penetration of the hole wavefunction tail further into the (Zn,Mn)Se matrix. The proposed model is supported by microscopic calculations which accurately include the role of hole-hole Coulomb interactions as well as the hole-Mn spin exchange interactions.

## I. INTRODUCTION

Unlike metals, semiconductors have relatively low carrier densities that can be drastically changed by either doping or by applying a perturbation, such as gate voltage. Magnetic semiconductors, show important and potentially useful differences [1-3] compared to ferromagnetic metals. For example, if the magnetism in a magnetically doped semiconductor is mediated by carriers, changes in the carrier density may significantly alter the exchange interaction between the carrier spins and those of the magnetic impurities. By changing the number of carriers using light excitation or by applying a gate voltage, it is possible to modify the effective strength of exchange coupling. When this effect is sufficient to turn ferromagnetism *on* or *off*, there arise intriguing possibilities for light- or bias-controlled ferromagnetism [4, 5], not possible in conventional ferromagnetic metals [6]. Even an increase in temperature, which leads to a higher carrier density, can enhance exchange coupling between magnetic impurities and result in the onset of ferromagnetism at elevated temperatures [7]. Furthermore, such a temperature-enhanced magnetic ordering is not limited to a mean-field picture and persist even when the fluctuations are included [8].

Epitaxially grown and colloidal semiconductor quantum dots allow for a versatile control of the number of carriers, their spins, and the quantum confinement which could lead to improved optical, transport, and magnetic properties as compared to their bulk counterparts [9-14]. In contrast to bulk structures, adding a single carrier in a magnetic QD can have important ramifications. An extra carrier can both strongly change the total carrier spin and the onset of magnetization [14, 15] that can be further controlled by modifying the quantum confinement and the strength of Coulomb interactions [16, 17]. There is encouraging experimental progress in Mn-doped QDs, from the controlled inclusion of a single Mn impurity to the onset of magnetization at temperatures substantially higher than their bulk counterparts [18-29].

Unlike the majority of experiments, carried out in Type-I band alignment QDs [18-33] where both electrons and holes are confined in the QD, we studied ZnTe QDs embedded in a

(Zn,Mn)Se magnetic matrix with a Type-II band alignment where only holes are confined [34]. The PL in this system is due to the recombination of holes confined in the ZnTe QDs with electrons which move outside the QD in the (Zn,Mn)Se magnetic matrix held by the electron-hole Coulomb attraction. In the presence of a magnetic field applied perpendicular to the QD plane, the PL peak energy exhibits a red shift and the PL becomes strongly circularly polarized. Both the peak energy and the circular polarization dependence on magnetic field follow a Brillouin-like behavior. Magnetic polarons (MPs) have been observed in this system for lower power excitation with 3.06 eV excitation (for which we have single hole occupancy) [34]. It was found that the MP formation energy decreases with increasing temperature and increasing magnetic field in agreement with results from similar systems [33]. The magnetic effects in our samples are due to the exchange interaction between the spins of the electrons and holes with the Mn spins (ferromagnetic interaction for electrons and antiferromagnetic interaction for the holes) [1]. In this work, we demonstrate optical control of magnetic interaction by changing the hole wavefunction overlap with the magnetic impurities in the matrix using two different energies of the exciting laser photons. By changing the photon energy, we observe an up to three-fold increase of the PL saturation red shift induced by an external magnetic field. This effect can be viewed as the optical analog of the gate-controlled overlap between the carrier wavefunction and magnetic impurities in a quantum well [35, 36]. In the theoretical description, we included hole-hole Coulomb interactions, often neglected in dilute magnetic semiconductors [1-3]; these interactions push the tail of the hole wavefunction from the ZnTe QD into the Mn-doped matrix. Further extension of the hole wavefunction is observed when the hole-Mn exchange is taken into account. The increased Mn-hole overlap is responsible for the observed higher PL saturation red shift.

## II. EXPERIMENTAL

We have used two samples grown on GaAs substrates by molecular beam epitaxy (MBE) in this study [37]. The growth sequence using migration enhanced epitaxy was as follows: deposition of several ZnSe monolayers; ZnSe buffer layer; a 50 nm (Zn,Mn)Se buffer layer; a ZnTe QD layer; a 50 nm (Zn,Mn)Se capping layer. The ZnTe QDs are disk shaped with average diameter of 20 nm and height of 3 nm. The two samples used in this work differ only in the composition of the (Zn,Mn)Se matrices. Sample 1 (sample 2) has a 10% (6%) Mn concentration. The samples were placed inside a 7 tesla optical magnet cryostat which operates in the 7-300 K temperature range. The experiments were carried out in the Faraday geometry with the emitted light propagating along the direction of the magnetic field applied along the normal to the QD plane. Either 3.06 eV ($\lambda$ = 405 nm) or 2.54 eV ($\lambda$ = 488 nm) photons were used to excite the photoluminescence. Both laser lines were linearly polarized. The PL was collected and focused onto the entrance slit of a single monochromator and detected by a cooled multichannel charged coupled device (CCD) detector. A combination of quarter wave plate and linear analyzer was placed just before the spectrometer entrance slit to separate the two circularly polarized components ($\sigma_+$ and $\sigma_-$) of the luminescence.

## III. RESULTS

In Fig. 1(a) and Fig. 1(b) we show a schematic of the band diagram in ZnTe QDs embedded in a (Zn,Mn)Se matrix [37]. The band alignment is Type-II with the holes confined in the ZnTe QDs and the electrons orbiting around the QDs inside the (Zn,Mn)Se matrix as shown in Fig. 1(c). The magnetic effects in our samples originate from the exchange interactions of electron and hole spins with those of the Mn ions. The holes even though confined in the ZnTe QDs interact with the Mn ions through the hole wavefunction tail that penetrates into the (Zn,Mn)Se magnetic matrix. We point out that the exchange constant $\beta$ between holes and Mn spins is 4-5 times larger than the corresponding exchange constant $\alpha$ that describes the electron-Mn spin interaction [1].

The circular polarization P of the PL [P is defined as the ratio $(I_+ - I_-)/(I_+ + I_-)$, where $I_+ (I_-)$ is the intensity of the $\sigma_+ (\sigma_-)$ PL component.] under 3.06 eV excitation from sample 1 is plotted as function of magnetic field B in Fig. 2. The polarization increases monotonically with B and saturates at approximately B = 3 tesla. The sign of the polarization indicates that the holes (electrons) participating in the recombination are in the $+3/2(-1/2)$ spin state. In contrast to the PL red shift, there is no significant dependence of the saturation value of P on the exciting laser photon energy.

In Fig. 3(a), we show the evolution of the PL from sample 1 with magnetic field at T = 7 K using 3.06 eV excitation. The peak energy was determined using an asymmetric Gaussian fitting and is indicated by the magenta open squares on each spectrum. The red shift $|\Delta E|$ at B = 7 T is equal to 5.3 meV. A similar plot for 2.54 eV excitation is shown in Fig. 3(b) with a red shift $|\Delta E|$ of 14.6 meV at B = 7 T. In this case, the fitted peak energy is indicated by blue open circles.

In Fig. 4 we plot the peak energies as function of magnetic field for sample 1 [10% Mn composition of the (Zn,Mn)Se matrix] for both excitation energies (magenta open square for 3.06 eV excitation, blue open circle for 2.54 eV excitation). It is clear that the red shift under 2.54 eV excitation is approximately 3 times the corresponding value for 3.06 eV excitation. In Fig. 5 we give the same plot for sample 2 [6% Mn composition for the (Zn,Mn)Se matrix]. The increase of the red shift under 2.54 eV excitation is present but less pronounced (red shift |ΔE| of 5.2 meV for 3.06 eV excitation and 6.9 meV for 2.54 eV excitation).

## IV. DISCUSSION

In this section we present a model with which we explain the difference in the red shifts under different laser photon energies. In Fig. 1(a) we show a schematic of the photo-excitation mechanisms for 3.06 eV. The 3.06 eV photon energy is above the bandgaps of the (Zn,Mn)Se matrix and the ZnTe QDs. Given that the volume of (Zn,Mn)Se top and bottom layers greatly exceeds the QD total volume, we can safely assume that under 3.06 eV excitation, most of the electron-hole pairs are created in the (Zn,Mn)Se matrix. Following photoexcitation, the electrons stay in the (Zn,Mn)Se matrix, while the holes which participate in the recombination at 2.05 eV are captured by the ZnTe QDs as shown in Fig. 1(a) by the dashed arrow. Once a hole is captured, it creates a repulsive Coulomb potential that prevents the capture of a second hole. In contrast, the 2.54 eV photon energy lies below the (Zn,Mn)Se bandgap and excites electron-hole pairs directly in the ZnTe QDs as shown in Fig. 1(b). Under these conditions, we can have multiple occupancy of holes in the ZnTe QDs. This is supported by the spectra shown in Fig. 6. The PL spectrum excited using 2.54 eV excitation is blue shifted by 18.9 meV with respect to the spectrum under 3.06 eV excitation. This has been observed previously in Type-II QDs [38, 39] and has been attributed to the Coulomb charging and the QD state filling of the holes. The Coulomb repulsion between holes occupying the same QD results in an extension of the hole wavefunction in the magnetic matrix which leads to an increase of the hole-Mn spin exchange energy; this results in the observed increase in the PL red shift with magnetic field. This model is in agreement with the results of a theoretical calculation which is discussed below.

We consider a QD occupied by two heavy holes and use a simple single-band model for two heavy holes of charge $e$ and effective mass $m^*$ confined in the QD plane ($x$-$y$ plane) by a parabolic potential described by an angular frequency $\omega_0$. We assume that a magnetic field $B$ perpendicular to the QD plane is applied. The Hamiltonian of the two-hole system confined in ZnTe QDs embedded in a nonmagnetic ZnSe matrix is the sum of the kinetic energy, the harmonic confinement and the Coulomb interaction between the two holes:

$$H_0 = \sum_{i=1,2} \left\{ \frac{\left[\mathbf{p}_i - (e/c)\mathbf{A}(\mathbf{r}_i)\right]^2}{2m^*} + \frac{1}{2}m^*\omega_0^2 r_i^2 \right\} + \frac{1}{4\pi\varepsilon}\frac{e^2}{|\mathbf{r}_1 - \mathbf{r}_2|}, \qquad (1)$$

here the holes are labeled by the index $i$; $\mathbf{p}_i = -i\hbar\nabla_i$ is the hole momentum operator; $c$ is the speed of light in vacuum; $\mathbf{A}(\mathbf{r}_i) = \frac{1}{2}(-By_i, -Bx_i)$ is the vector potential defined in the symmetric gauge; $\mathbf{r}_i = (x_i, y_i)$ is the two dimensional position vector operator; and $\varepsilon$ is the dielectric constant of the QD. The bulk parameters for ZnTe are $m^* = 0.2$ and $\varepsilon = 9.4$ [40]. The ZnTe QD width is defined as $l_0 = \sqrt{\hbar/m^*\omega_0}$ while the classical turning radius of this oscillator is expressed as $R_{cls} = \sqrt{2}l_0$. In our model we take the cylindrical surface of radius $R_{cls}$ and height $h_{QD}$ to be the boundary between the QD and the surrounding matrix. Our disk shaped QDs have a diameter greater than the height $h_{QD}$; as a result, there is a high anisotropy between $z$ and $x$-$y$ directions for heavy holes due to the spin-orbit interaction [27, 41, 42]. We now consider a system with magnetic (Zn,Mn)Se matrix. The exchange interaction between the heavy hole spins and $N_{Mn}$ manganese spins located outside the QD is described by an Ising-like Hamiltonian [43, 44],

$$H_{ex} = -\frac{\beta}{3}\sum_{i=1,2}\sum_{j=1}^{N_{Mn}} J_{zi}S_{zj}\delta(\mathbf{r}_i - \mathbf{R}_j), \qquad (2)$$

where $\beta$ is the hole-Mn exchange coupling constant, $j$ is the Mn index, $J_{zi}$ is the $z$- component of the $i$'th heavy hole (pseudo) spin operator ($J = 3/2$, $J_{zi} = \pm 3/2$), $S_{zj}$ is the z-component the $j$'th Mn ion spin operator ($S = 5/2$) and $\mathbf{R}_j$ is the position operator for the Mn ion. The p-d exchange integral for (Zn,Mn)Se $N_0\beta = 1.11$ eV [1]. We consider the saturated regime ($T = 0$ K), where the $j$'th Mn spins are fully spin polarized and continuously distributed outside the QD ($R \geq R_{\text{cls}}$). Therefore, the calculation is relevant for the high field limit ($B > 4$ T). The sum over the Mn ions in Eq. (2) can now be approximated by an integral, $\sum_j S_{zj} \to N_0 x_{\text{Mn}} \int d^3 R \, S(R)$, where $N_0$ is the cation density, $x_{\text{Mn}}$ is the effective Mn molar fraction and $S(R)$ is the Mn spin profile; $S(R) = -5/2$ for $R > R_{\text{cls}}$ and $S(R) = 0$ for $R < R_{\text{cls}}$. We note that, $x_{\text{Mn}}$ is related to the Mn molar fraction $x$ by: $x_{\text{Mn}} = x(1-x)^{12}$ [45]. The total Hamiltonian $H$ is: $H = H_0 + H_{\text{ex}}$.

We diagonalize $H_0$ using the many-body basis and obtain the energies and wavefunctions of the heavy holes. The Hamiltonian matrix is constructed in the basis of Slater determinants of the single-particle (2D harmonic oscillator) wavefunctions normalized over the QD height. All matrix elements of $H_0$ are calculated analytically. To simulate the strong confinement of the holes in the dot, we consider a QD with a smaller classical radius than the average QD radius. The hole localization diameter can be smaller than the QD diameter due to alloy and spin disorder scattering [34, 46, 47]. We assume a QD with an in-plane confinement of $\hbar\omega_0 = 15.5$ meV, corresponding to $R_{\text{cls}} = 7$ nm of height $h_{\text{QD}} = 3$ nm in an external magnetic field $B = 7$ T. A schematic of the spin alignment is given in Fig. 7. The red (black) arrows represent the hole (electron) spins; the applied magnetic field symbol is the green arrow; the Mn spins are indicated by the orange arrows. From the polarization data of Fig. 2, two heavy holes occupying the ZnTe QD are in their triplet state. The Mn ions in the magnetic matrix (outside the dot $R \geq R_{\text{cls}}$) have their spins pointing in a direction opposite to that of the hole spins. In contrast, the electron spins are parallel to the spins of the Mn ions.

**Non-magnetic matrix**: To explore the effect that the hole-hole Coulomb interaction has on the hole wavefunction, we first consider the QD embedded in a nonmagnetic matrix ($x_{Mn} = 0$). The carrier density as a function of $r$ is shown in Fig. 8(a) without Coulomb interaction (black, solid line) and with Coulomb interaction (red, dotted line). A hole density contour plot is illustrated in Fig. 8(b) when there is no Coulomb interaction. In Fig. 8(c) we show the same plot with the Coulomb interaction included. In the absence of this interaction, the holes are confined closer the QD center [Fig. 8(a) solid line and Fig. 8(b)] with a 24% probability that the two heavy holes are outside $R_{cls}$. When Coulomb interaction is included in the model [Fig. 8(a) dotted line and Fig. 8(c)], there is a 32% probability that the holes are outside $R_{cls}$. In simple terms, the two holes are pushed away from the QD center towards $R_{cls}$ due to their mutual repulsion. The increase in the carrier density outside $R_{cls}$ confirms that the Coulomb repulsion between holes results in an extension of their wavefunction outside $R_{cls}$ and into the (Zn,Mn)Se matrix.

**Magnetic matrix**: We next turn to a QD with a magnetic matrix ($x_{Mn} \neq 0$). Figure 9(a) shows a plot of the hole density as a function of $r$ for the effective Mn molar fraction $x_{Mn}$ of 1% ($x = 1\%$, red, solid line), 2% ($x = 3\%$, blue, dot-dashed line) and 3% ($x = 8\%$, violet, dotted line). From Fig. 9(a), it can be seen that as the $x_{Mn}$ increases, the hole density maximum moves from near the center of the QD to a position beyond $R_{cls}$ [the QD/(Zn,Mn)Se matrix boundary]. This is further illustrated in the density plots of Figs. 9(b), 9(c) and 9(d). We find that the probability of finding the holes outside $R_{cls}$ is 45%, 61% and 74% for $x_{Mn}$ of 1%, 2% and 3% respectively. The additional shift of the hole density maximum with increasing $x$ results from the fact that energetically it is more favorable for the heavy holes to increase their overlap with the Mn outside the QD. The exchange energy increase with the additional overlap of the hole spin density with the Mn ions is responsible for the corresponding increase of the PL red shift shown in Figs. 4 and 5 for sample 1 and 2, respectively.

## V. CONCLUSIONS

We have observed a strong dependence of the PL energy saturation red shift with magnetic field on excitation photon energy in ZnTe QDs embedded in a (Zn,Mn)Se matrix. For photon energies below the (Zn,Mn)Se gap (above the ZnTe gap), we measured an increase of the magnetic red shift up to three times compared to the red shift of PL excited with photons having energy above the matrix bandgap. These observations are interpreted as follows: exciting photons with energy above the matrix gap (2.8 eV) result in single hole occupancy of the ZnTe QDs; excitation with photon energy below the (Zn,Mn)Se matrix gap on the other hand, leads to multiple hole occupancy. The hole-hole Coulomb repulsion result in an extension of the hole wavefunction further into the (Zn,Mn)Se matrix and thus an increase of the hole-Mn exchange energy. This model is supported by a calculation of the hole wavefunction. Our system is one in which we have optical control of carrier wavefunctions.

## Acknowledgements:

This work was supported by NSF DMR-1305770 and US DOE, Basic Energy Sciences, under Award DE-SC0004890.

**Fig. 1**

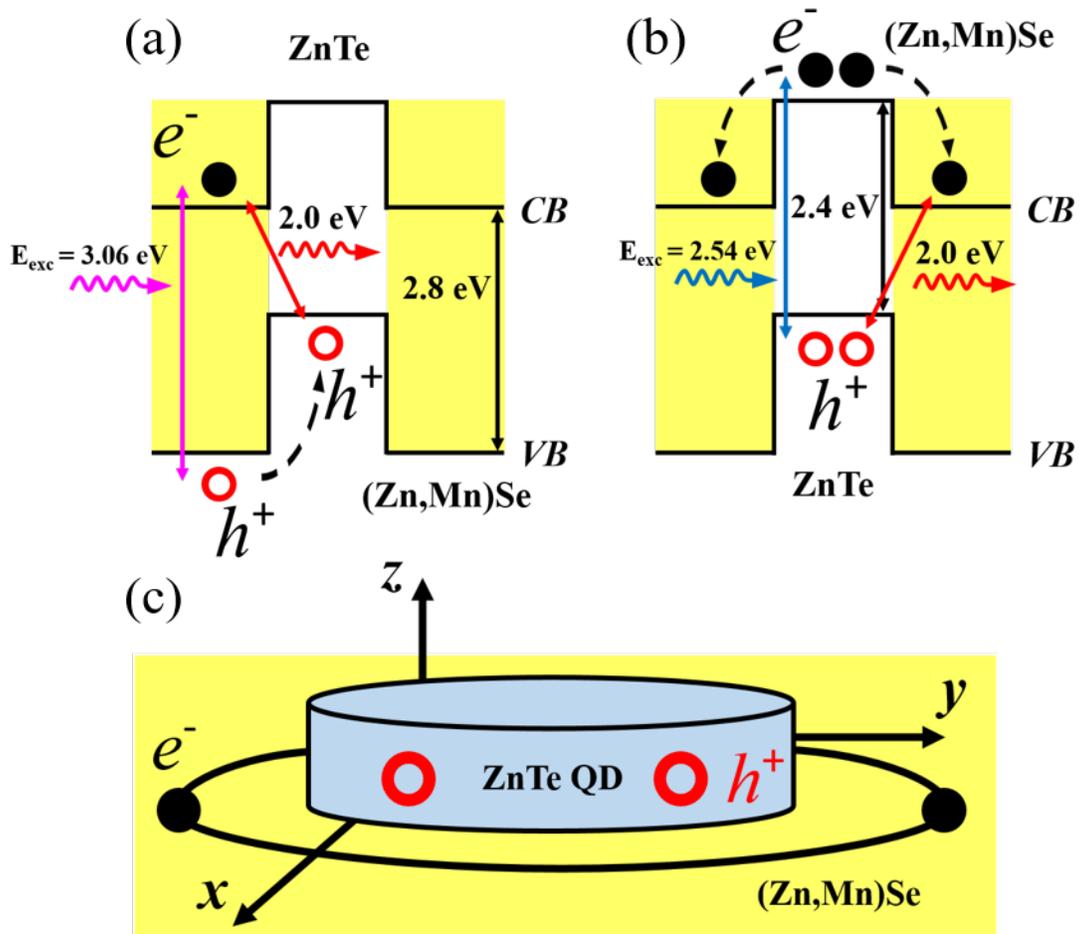

**Fig. 1:** Band structure diagram of a ZnTe QD embedded in a (Zn,Mn)Se matrix that shows the PL excitation using (a) 3.06 eV and (b) 2.54 eV photons. (c) Schematic of a flat disc-shaped ZnTe QD. The yellow background indicates the (Zn,Mn)Se matrix.

**Fig. 2**

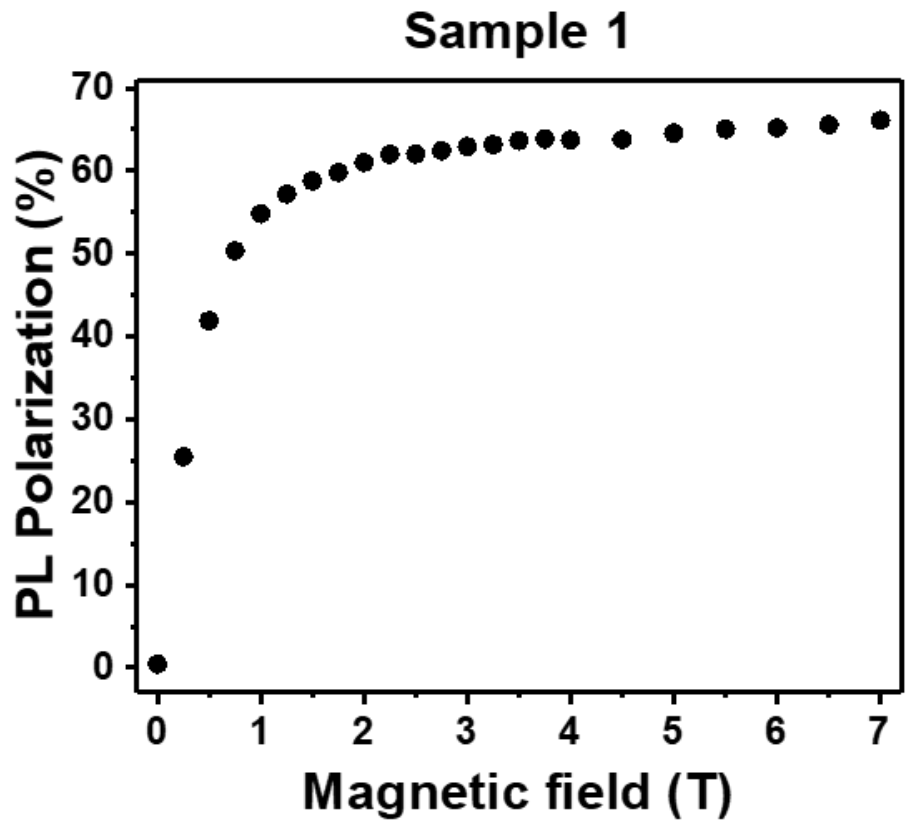

**Fig. 2:** The PL circular polarization from sample 1 as function of applied magnetic field at $T = 7$ K for photon excitation energy of 3.06 eV.

**Fig. 3**

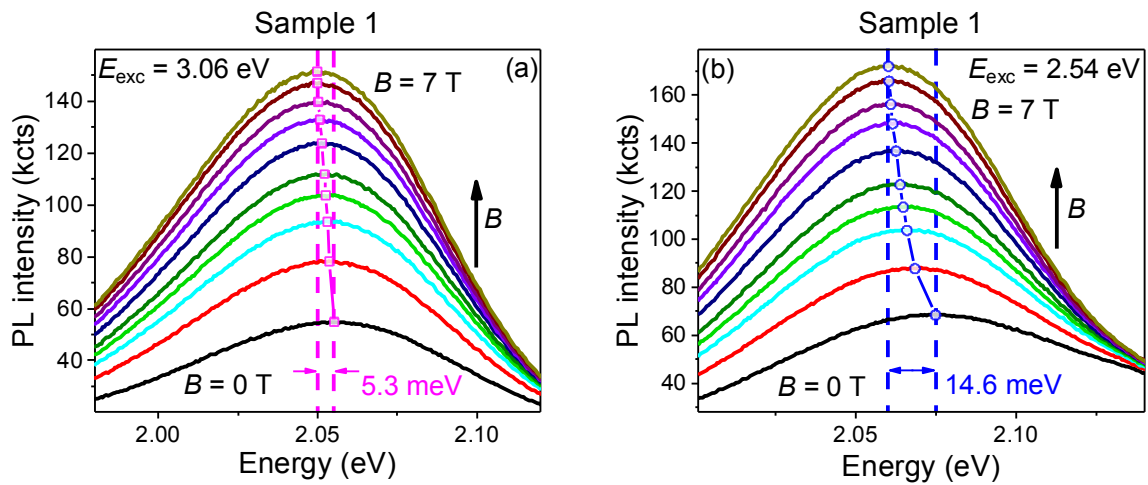

Fig. 3: Evolution of the PL spectra from sample 1 with applied magnetic field. Photon excitation energy: (a) 3.06 eV and (b) 2.54 eV. The squares and circles in (a) and (b) represent the fitted PL peak energies. The magnetic field values are: 0, 0.5, 1, 1.5, 2, 3, 4, 5, 6, 7 T.

**Fig. 4**

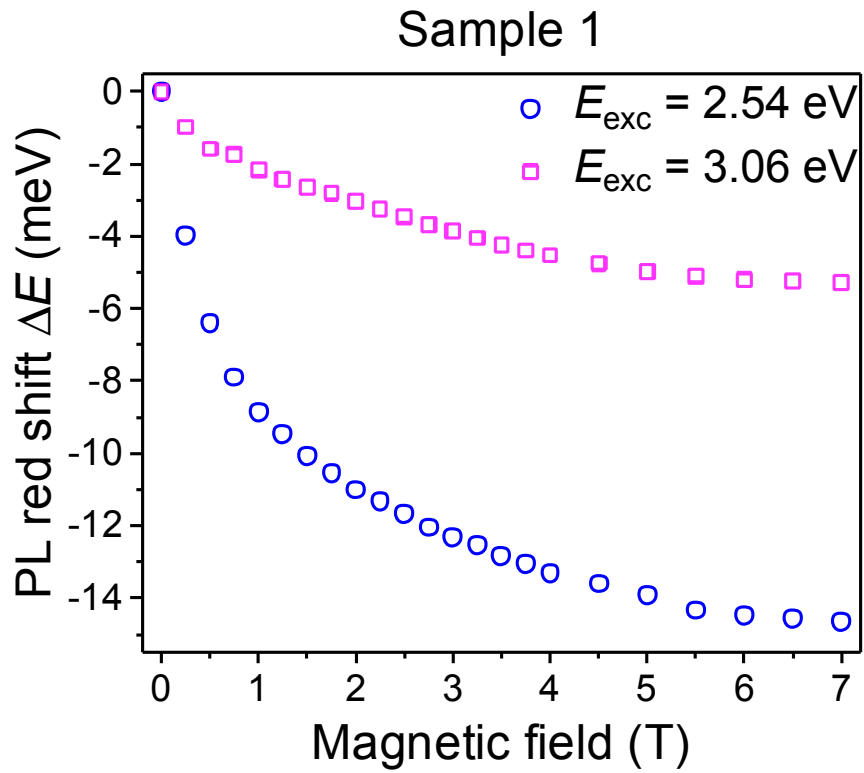

Fig. 4: Fitted PL peak energy relative to the peak energy at zero field, plotted as function of magnetic field for sample 1 at $T = 7$ K. Photon excitation energy: 3.06 eV (squares) and 2.54 eV (circles).

**Fig. 5**

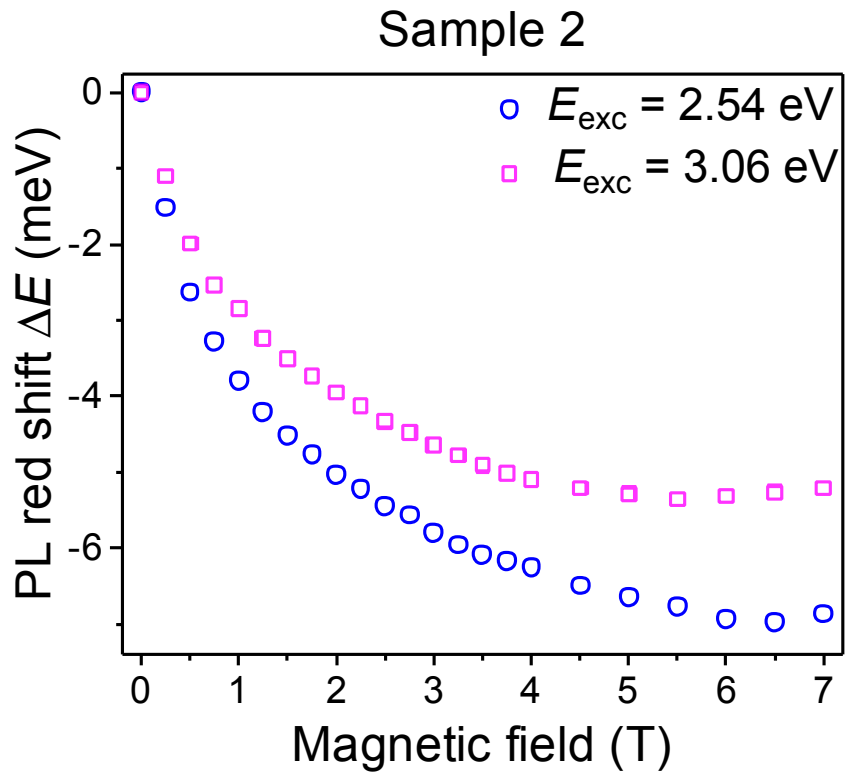

Fig. 5: Fitted PL peak energy relative to the peak energy at zero field, plotted as function of magnetic field for sample 2 at $T = 7$ K. Photon excitation energy: 3.06 eV (squares) and 2.54 eV (circles).

Fig. 6

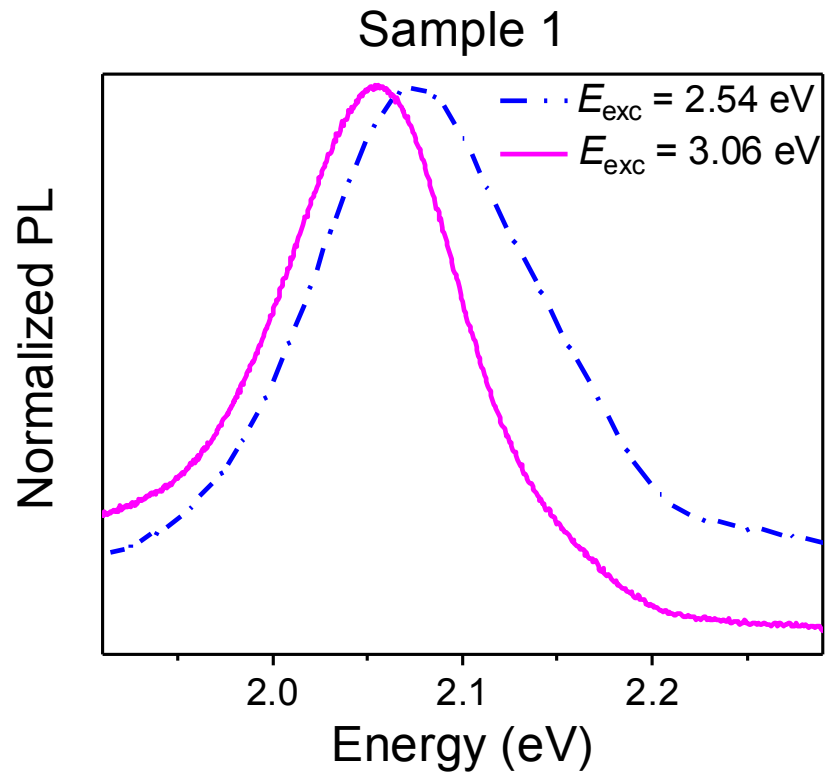

Fig. 6: Zero field PL spectra from sample 1 at $T = 7$ K. Photon excitation energy: 3.06 eV (solid line) and 2.54 eV (dot-dashed line).

**Fig. 7**

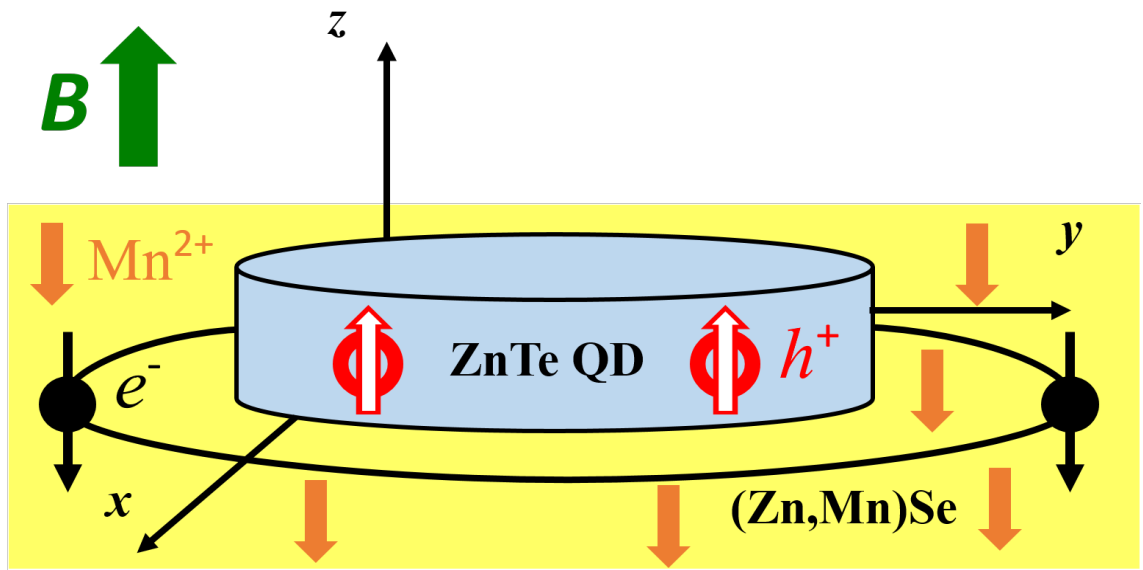

Fig. 7: Schematic of the electron, hole, and Mn ion spin orientation in the presence of a magnetic field *B* (green arrows). Electron spins are denoted by black arrows, hole spins by red arrows and Mn ion spins by orange arrows.

**Fig. 8**

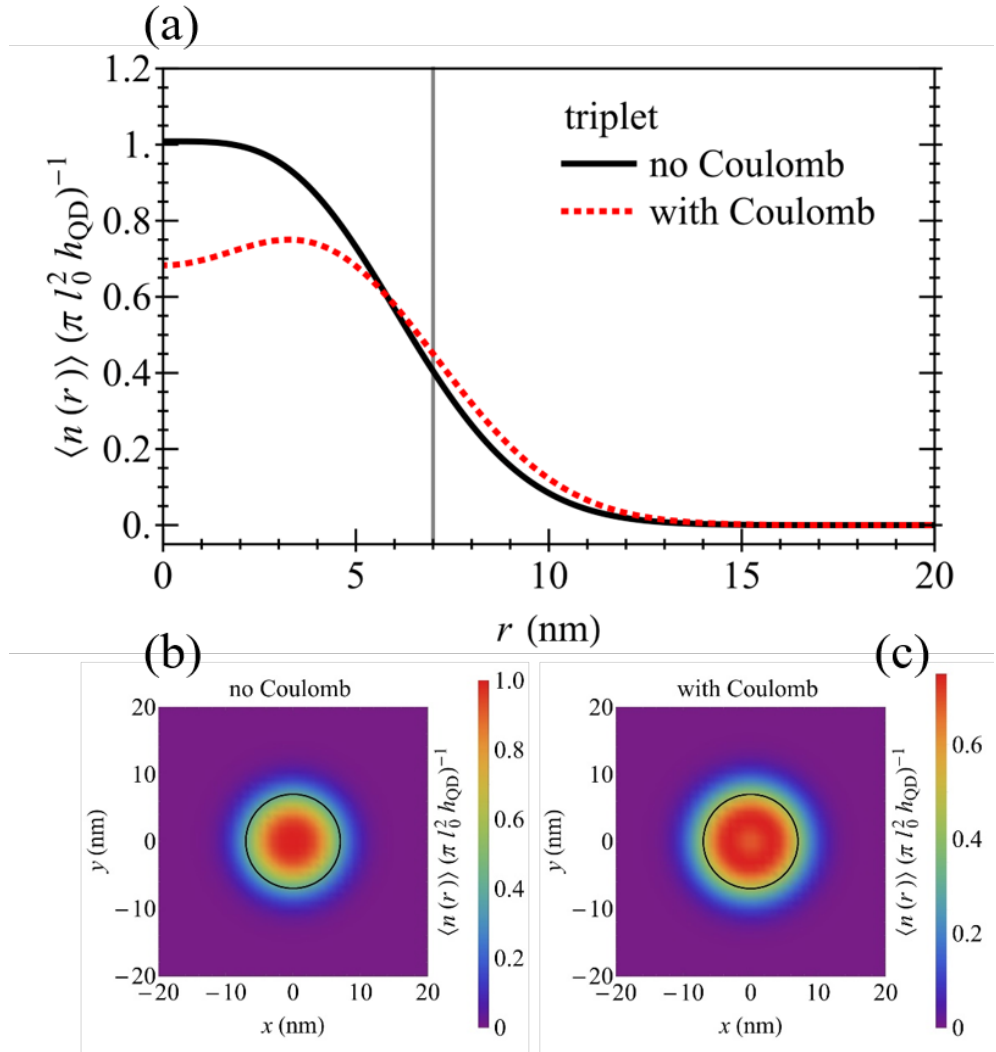

Fig. 8: (a) Calculated hole density as function of distance $r$ from the QD center at $B = 7$ T. Non-magnetic ZnSe matrix is assumed. The results are calculated without (black solid line) and with (red dotted line) the inclusion of the hole-hole Coulomb repulsion. Contour plots for the hole density without (b) and with (c) the inclusion of the hole-hole Coulomb repulsion. Red (purple) indicates high (low) hole density. The black circles in (b) and (c) represent $R_{cls}$.

**Fig. 9**

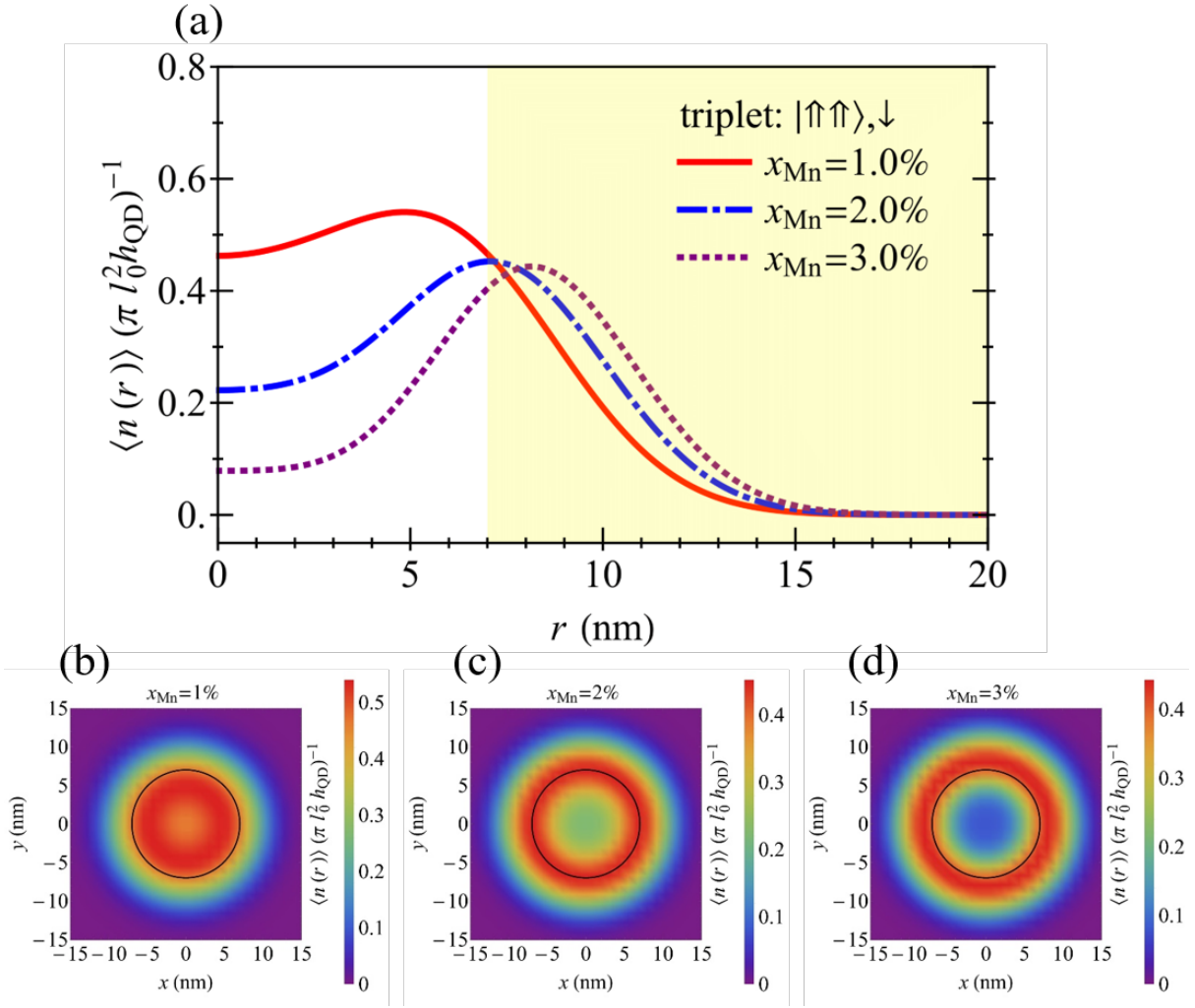

**Fig. 9:** (a) Calculated hole density as function of distance $r$ from the QD center at $B = 7$ T assuming magnetic (Zn,Mn)Se matrices with the inclusion of the hole-hole Coulomb repulsion. Red solid line: $x_{Mn} = 1\%$ ($x = 1\%$); blue, dot-dashed line: $x_{Mn} = 2\%$ ($x = 3\%$); violet dotted line: $x_{Mn} = 3\%$ ($x = 8\%$). Contour plots for the hole density: (b) $x_{Mn} = 1\%$; (c) $x_{Mn} = 2\%$; (d) $x_{Mn} = 3\%$. Red (purple) indicates high (low) hole density. The black circles in (b), (c) and (d) represent $R_{cls}$.